\documentclass[12pt,preprint]{aastex}
\usepackage{natbib,epsfig}
\tighten

\newcommand{\z}{{\it z}}
\newcommand{\zre}{{\it z$_{reion}$}}
\newcommand{\sst}{{\it Spitzer}}

\newcommand{\mun}{M$_{\sun}$~Mpc$^{-3}$}
\newcommand{\lun}{L$_{\sun}$~Mpc$^{-3}$}

\slugcomment{To appear in the {\it Astrophysical Journal}}
\shorttitle{High Redshift Stellar IMF}
\shortauthors{Chary}

\begin{document}

\title{The Stellar Initial Mass Function at the Epoch of Reionization}

\author{Ranga-Ram Chary\altaffilmark{1}}
\altaffiltext{1}{{\it Spitzer} Science Center MS220-6, California Institute of 
Technology, Pasadena, CA 91125; {\tt rchary@caltech.edu}}

\begin{abstract}

I provide estimates of the ultraviolet and visible light luminosity density at $z\sim6$
after accounting for the contribution from
faint galaxies below the detection limit of deep {\it Hubble} and {\it Spitzer}
surveys. I find the rest-frame $V-$band luminosity density is a factor
of $\sim2-3$ below the ultraviolet luminosity density at $z\sim$6. This implies that the
maximal
age of the stellar population at $z\sim6$, for a Salpeter initial mass function, and a single,
passively evolving burst, must be $\lesssim$100 Myr.
If the stars in $z\sim6$ galaxies are remnants of the star-formation
that was responsible for ionizing the intergalactic medium, reionization must have been a brief
process that was completed at $z<7$. This
assumes the most current estimates of the clumping factor and escape fraction and
a Salpeter slope extending up to 200 M$_{\sun}$ for the stellar initial mass 
function (IMF; dN/dM$\propto$M$^{\alpha}$, $\alpha=-2.3$).
Unless the ratio of the clumping factor to escape fraction is less than 60,
a Salpeter slope for the stellar IMF and reionization redshift higher than 7 is ruled out. In order to maintain an
ionized intergalactic medium from redshift 9 onwards, the stellar IMF must have a 
slope of $\alpha=-1.65$ even if stars as massive as $\sim$200\,M$_{\sun}$ are formed. Correspondingly,
if the intergalactic medium was ionized from redshift 11 onwards, the IMF must have  $\alpha\sim-1.5$.
The range of stellar mass densities at $z\sim6$ straddled by IMFs which result in reionization at $z>7$
is 1.3$\pm$0.4$\times$10$^{7}$~\mun. 

\end{abstract}


\section{Introduction}

Deep optical/infrared imaging surveys with {\it Hubble} and {\it Spitzer} are providing an unprecedented
opportunity to study star-forming galaxies at high redshifts. It is increasingly clear that these
star-forming galaxies are responsible for reionizing the intergalactic medium (IGM), a process which seems to
complete by $z\sim6$ \citep{Becker:01, Fan:06}. The shape 
of the luminosity function of
high redshift active galactic nuclei detected by the Sloan Digital Sky Survey (SDSS) indicates that ionizing
photons from supermassive
black holes fall many orders of magnitude short of the minimum required to ionize the IGM \citep{Fan:01}. 
Even the current star-formation in Lyman-break galaxies (LBGs) at $z\sim6$ as measured by \citet{Bouwens06, Bouwens:07},
falls about a factor of $6-9$ below the minimum required
to maintain an ionized IGM for canonical estimates of two
key parameters, the clumping factor of the gas and the escape fraction, assuming a Salpeter initial mass function (IMF). 
One possibility is that the contribution from star-formation
in faint galaxies, below the detection threshold of current surveys, is higher than has been estimated. This
translates to a faint end slope for the galaxy ultraviolet luminosity function of $-1.9$, a value
that is considerably steeper than the value of $-1.74$ that has currently been measured \citep{Bouwens:07}. Evidence
for such a steep slope is present in estimates of star-formation rate densities derived from the comoving gamma-ray 
burst number density \citep{Chary:07a}. However, that measurement
has large uncertainties limited by the small number of high redshift GRBs currently known and the uncertain conversion
between GRB number density and star-formation rate density. 

An alternate possibility is that the star-formation rate in galaxies currently detected at $z\sim6$ was higher
in the past. This implies that the ultraviolet luminosity function of galaxies must show luminosity evolution
between $z\sim6$ and higher redshifts. Although constraints on the ultraviolet luminosity function at
higher redshift are difficult to obtain from existing data, the upper limits on bright Lyman-break galaxies 
at $z>>6$, seem to rule out such an evolution \citep{Bouwens:04}. In fact, the upper limits seem to suggest a negative
evolution with increasing redshift, in the sense that galaxies
at the bright end of the UV luminosity function at $z\sim6$ were fainter at higher redshift in the ultraviolet. There
are no available constraints on the evolution of the faint end of the UV luminosity function at $z>6$. 

Measurement of the stellar mass density provides an alternate constraint
on the past history of star-formation. Deep
{\it Spitzer} surveys which detect the rest-frame visible light from high redshift galaxies allow the stellar
mass in galaxies at $z\sim6$ to be measured \citep{Yan:06, Stark:06, Eyles:05}. The galaxies which 
are detected in rest-frame optical light, are primarily the brightest
galaxies in the ultraviolet. The stellar mass density directly inferred from the {\it Spitzer}
detections and stacking analysis are therefore, a lower limit. 
By making reasonable assumptions for the rest-frame $V-$band luminosity of galaxies which are undetected in the {\it Spitzer}
data but seen in the ultraviolet,
it is possible to estimate the true co-moving stellar mass density at $z\sim6$. 
Since the stars at $z\sim6$ are remnants of 
star-formation at higher redshifts, we can use this stellar mass density estimate to
constrain the number of ionizing photons 
produced by star-formation in the past
for comparison with the minimum number of photons required to keep the IGM ionized.

At present, the exact redshift of reionization is not known.  {\it Wilkinson} Microwave Anisotropy
Probe (WMAP) polarization measurements indicate
a large optical depth to scattering of microwave background photons suggesting that the  
intergalactic medium (IGM) was ionized 
at some time between redshifts of 7 and 14 \citep{Kogut:03, Spergel:07}. 

In this paper, we 
estimate the optical and ultraviolet luminosity density at $z\sim6$ after correcting for faint galaxies
undetected by current surveys. We then calculate the stellar mass density and average age of the stellar population
in galaxies,
required to reproduce the luminosity density estimates for different shapes of the stellar IMF. We also quantify
the number of ionizing photons produced over the past history of the starburst
and assess if they are sufficient to ionize the IGM at $z>6$.
Finally, we constrain the stellar initial mass function at $z>6$ as a function of the
reionization history of the Universe.
A standard $\Omega_{\rm M}$=0.27, $\Omega_{\Lambda}$=0.73, H$_{\rm 0}$=71~km~s$^{-1}$~Mpc$^{-1}$ cosmology 
is adopted throughout this paper.

\section{Optical and Ultraviolet Luminosity Density at \z$\sim$6}

Deep \sst\ imaging data which detects the rest-frame visible emission of $z>6$ galaxies has yielded the first constraints
on the stellar mass density at $z\sim6$. \citet{Yan:06} and \citet{Eyles:05} measured the IRAC 
photometry of Lyman-break galaxies selected
using the $i-$dropout technique in the Great Observatories Origins Deep Survey (GOODS) fields. By fitting the multiband
photometry of each galaxy with population synthesis models and deriving the stellar mass of each galaxy, they
find a lower limit to the stellar mass density of $0.1-0.7\times10^{7}$~\mun,
assuming a Salpeter initial mass function. 
The value is a lower limit since only the brightest galaxies in the ultraviolet are typically detected while many others
are blended with brighter foreground sources in the IRAC images.
At lower redshifts of $\z\sim5$, \citet{Stark:06} estimate the  stellar mass density to be $0.6-1\times10^{7}$~\mun\ by combining
a spectroscopic and photometric redshift sample in GOODS-S. \citet{Chary} adopt an alternate technique where they fit for 
the stellar masses of only a spectroscopic sample of galaxies at $5<z<6.5$ to derive a lower limit to the stellar mass density
of 0.3$-$1.1$\times10^{6}$~\mun\ at $z\sim6$. They adopt the ultraviolet luminosity function of \citet{Bouwens:07}
and make the assumption that the rest-frame 
$UV/V-$band luminosity ratios of the {\it Spitzer} undetected galaxies at these
redshifts are the same as the detected galaxies.
The {\it Spitzer} detected galaxies have $\nu L_{\nu}$(1500\AA)$\gtrsim2\times$10$^{10}$~L$_{\sun}$ and
$\nu L_{\nu}$(5500\AA)$\gtrsim10^{10}$~L$_{\sun}$. The median visible to ultraviolet luminosity ratio for
the \sst\ detected galaxies is 0.5.
They then apply a completeness correction by scaling the UV luminosity density of their spectroscopic
sample to the total UV luminosity density of Lyman-break galaxies at $z\sim6$ as estimated by \citet{Bouwens:07}.
This completeness correction is a factor of 34 which results in a total stellar mass density 
of $1-3.7\times10^{7}$~\mun\ at $z\sim6$ for a Salpeter IMF. The corresponding
UV and V-band luminosity densities derived by integrating 
down to a UV luminosity of 10$^{7}$~L$_{\sun}$ is 1.4$\times$10$^{8}$~\lun\ and
6.6$\times$10$^{7}$~\lun, respectively. We call this the ``high-V" case.
Integrating to zero luminosity results in only 
a 10\% upward correction to these values. These values are in excellent agreement with 
the stellar mass density at $z\sim6$ obtained in the hydrodynamic
simulations of \citet{Dave:06} but substantially lower than the value of 8.8$\times$10$^{7}$\,\mun\ derived
by \citet{Nagamine}.

The assumption that the visible to ultraviolet luminosity ratios
of faint LBGs is similar to that of brighter LBGs
is subject to some uncertainty. Faint LBGs are predominantly in low mass halos. The 
supernova feedback from the initial burst of star-formation might 
be sufficient to inhibit or delay further star-formation. This would
suggest that UV faint galaxies have lower visible to ultraviolet luminosity ratios
than galaxies at the bright end of the UV luminosity function. This 
trend between rest-frame optical-ultraviolet colors with ultraviolet luminosity
appears to be seen in semi-analytical 
models \citep[][personal communication]{Somerville}. To assess
the dependence of the $z\sim6$ luminosity density on the optical luminosities of faint galaxies, we 
assume that the $UV/V-$band luminosity ratios of faint {\it Spitzer} undetected galaxies follow the 
correponding ratios seen in semi-analytical models. The ratio of the visible to ultraviolet luminosity
density in the models, for galaxies below the {\it Spitzer} detection threshold, is 0.28. 
For the UV bright galaxies, some of which are
undetected due to blending, we adopt the $UV/V-$band luminosity ratios of the {\it Spitzer} detected sources
which are similar to the colors of UV bright galaxies in the semi-analytical models.
We find that the resultant V-band luminosity density under these assumptions, obtained by integrating 
down to $\nu L_{\nu}$(1350\AA)$\sim$10$^{7}$~L$_{\sun}$ is 4.2$\times$10$^{7}$~\lun\ which is 30\%
smaller than the estimate of \citet{Chary}. We call this the ``low-V" case.

More recently, massive, evolved stellar populations have been claimed in candidate high redshift galaxies
which would boost the derived $V-$band luminosity densities 
and therefore the stellar mass densities, by factors
of several \citep{Wiklind:07, Panagia:05}.  However, since a significant fraction of 
such sources are detected at 24$\mu$m, a wavelength which traces the 
dust emission from galaxies, the evidence is strongly against these objects being
at high redshifts \citep{Chary:07b}\footnote{The exception is a galaxy at z=5.554 which might have its 
stellar mass estimate erroneously elevated due to the presence of H$\alpha$ in the \sst\ passbands. See source
ID44 in Chary et al. (2007) for details.}. Furthermore, 
through mid-infrared photometric redshifts, it has been demonstrated that
the best candidate among these proposed high redshift galaxies is a dusty,
infrared luminous galaxies at $z\sim1.7$ \citep{Chary:07b}. We ignore this population of galaxies for the rest of the
discussion until there is any reliable evidence that they really are at $z>>5$.

Typically, stellar mass densities are estimated by fitting the multi-wavelength spectral energy distribution of 
individual galaxies with population synthesis models and integrating over the resultant stellar mass distribution.
However, the stellar mass density is sensitive to the choice of the stellar mass function,
which is one of the parameters we are trying to constrain. 
Furthermore, since we do not have multiwavelength detections of faint galaxies, the only available approach 
involves fitting a population synthesis model to the 
integrated rest-frame UV and V-band luminosity density estimates at $z\sim6$ derived above. 

We adopt the Starburst99 set of models
which have the Padova tracks for the thermally pulsing asymptotic giant branch stars (TP-AGB) incorporated
\citep{Leitherer:99, Vaz:05}. The variables in the models are the slope of the stellar IMF, metallicity,
age of the stellar population and stellar mass density.
We consider both 0.02$Z_{\sun}$ and 0.4$Z_{\sun}$ templates and single burst models. 
We also convolve the templates with a model for the absorption by the Ly$-\alpha$ forest \citep{Madau:95}.
The fit to the luminosity density constrains the stellar mass density as well as the total number
of ionizing photons produced over the duration of the starburst. 
Adopting a single burst model yields maximal ages 
and stellar mass densities while the total number of ionizing
photons produced is insensitive to the time history of star-formation.

If we adopt a Salpeter IMF with metallicity 0.02\,Z$_{\sun}$, we find that the optical and ultraviolet luminosity
density at $z\sim6$ are fit by a stellar population of age 100 Myr and with a stellar mass density
of 1.6$\times$10$^{7}$\,\mun\ (Figure 1).
If instead we consider the ``low-V" case,
we derive a maximal stellar age of 50 Myr and a total stellar 
mass of 0.8$\times$10$^{7}$\,\mun.
The presence of extinction at $z\sim6$ would imply that reproducing
the observed optical to ultraviolet luminosity densities will require a bluer intrinsic stellar 
population, corresponding to a younger age. Since the average extinction in star-forming regions at $z\sim6$ is unknown,
we do not attempt to obtain fits that include extinction.

Changing the stellar initial mass function will change the derived stellar mass densities. Since we use
the optical and ultraviolet luminosity densities at $z\sim6$ as the overall constraint, and evolve a single star
burst of fixed mass, the true stellar mass density is the normalization factor for the fixed mass burst. Thus, for every stellar
IMF that we consider, we can derive the age, stellar mass density and number of ionizing photons produced by requiring
that the luminosity density estimates at $z\sim6$ are reproduced. We note that the age is not a strongly constrained
parameter since we are fitting to only observations at two wavelengths and the exact age would strongly
depend on the average time history of star-formation. By adopting a single burst model, we only derive maximal ages
for a single stellar population.

The ultraviolet and optical luminosity density at $z\sim6$ are dominated by ongoing star-formation
and intermediate mass stars respectively, the latter depending on the age of the starburst. For a 
particular IMF, the stars with lifetimes shorter than the age of the starburst at
that redshift would have presumably evolved to their end states i.e. neutron stars or black holes. They would still
be contributing to the total stellar mass density in the models as well as the total number of
ionizing photons produced by the starburst, even if they dont contribute to the ultraviolet/optical luminosities at $z\sim6$.

An upper limit to the stellar mass density for each IMF can also be derived by adopting a two component
stellar population synthesis model (Figure 1, right panel). The young stellar 
population, which dominates the UV luminosity
density is assumed to be $\sim$10 Myr old. An old, high mass-to-light ratio
stellar population is assumed to form in a single
burst at $z\sim20$ and evolve passively to $z\sim6$. This implies a stellar age of 770 Myr.
Naturally, a stellar IMF in which all stars have lifetimes younger than 770 Myr will yield unphysical
stellar masses and is represented with an ellipsis in Table 2. We note that such a two component
starburst has been considered previously by \citet{Chary} who showed that the number of ionizing photons
produced even in such an extreme scenario falls short of maintaining
an ionized IGM, assuming a Salpeter IMF. 
Although concentrating all
the star-formation into a single $z\sim20$ burst increases the total number of ionizing photons produced
by factors of a few (see Table 1 and 2), the recombination rate increases as $(1+z)^3$ which implies
that a much larger number of ionizing photons are required to keep the Universe ionized at these redshifts.

\section{Constraining Reionization}

By fitting the measured optical and ultraviolet luminosity density at $z\sim6$ with a evolving, single burst model,
we can determine the age of the stellar population and the stellar mass density. Integration of the
number of photons produced shortward of the Lyman limit over the age of the population, yields the total number
of ionizing photons produced by a starburst template which fits the luminosity density values. 
The integral constraint has the
advantage of being independent of the exact star-formation history which is likely to have a multitude of ages and
e-folding times. Since the vast majority of ionizing photons are produced in massive stars, the single burst model
also has the advantage that the time integral of the number of ionizing photons, for starburst ages larger than $\sim$10 Myr,
converges to a single value.

The number of ionizing photons produced over the history of star-formation can be compared with the number of
ionizing photons required to ionize the IGM. 
The number of photons required to initiate reionization is only
1 photon per baryon. However, to account for recombination, the production rate of ionizing photons per baryon must be 
$\sim$10 times higher. To quantify the rate of ionizing photons $\dot{N}(z)$ required to keep the IGM ionized at
any redshift, the rate of ionizing photons must be greater than or equal to the recombination rate.

The recombination rate R is:
\begin{equation}
R=n_{{\rm e}}~n_{{\rm HII}}~\alpha_{B}~C~~{\rm s}^{-1} {\rm Mpc}^{-3}
\end{equation}
where $n_{{\rm e}}$ is the comoving electron 
density, $n_{{\rm HII}}$ is the comoving ionized hydrogen density, $\alpha_{B}$ is the recombination coefficient
to excited states of hydrogen while C is the ionized hydrogen clumping factor defined as 
$<n_{{\rm HII}}^2>$/$<n_{{\rm HII}}>^2$. Since HeII reionization takes place at much lower redshifts $z\sim3-4$
\citep{Sokasian:02}, the comoving electron density is:
\begin{equation}
n_{\rm e} = n_{{\rm HII}} + n_{{\rm HeII}}
\end{equation}
and not $n_{{\rm HII}}$+2$n_{{\rm He}}$ as has previously been adopted, although the difference is negligibly
small because of the small fractional number density of He.
We adopt $\alpha_{B}$=2.59$\times10^{-13}(1+z)^3$~cm$^{3}$~s$^{-1}$ for the comoving recombination coefficient (Case B), assuming
a gas temperature of 10$^{4}$\,K. So, the photon production rate per baryon to balance the recombination rate
at any redshift, is:
\begin{equation}
\dot{N}(z) = \frac{R}{n_{\rm b}}~~s^{-1}
\end{equation}
The number of photons per baryon to start reionization at any redshift $N_0(z)$ is simply the ionized hydrogen fraction at
that redshift.
\begin{equation}
N_0(z) = \frac{n_{\rm HII}}{n_{\rm b}}
\end{equation}
So, the total number of ionizing photons per baryon
required to maintain the increasing ionized fraction of the IGM between any redshift $z_0$
and redshift $z$, where $z_0>z$, is:
\begin{equation}
N(z) = N_0(z) + \int_{z'=z_0}^{z'=z} \dot{N}(z') \frac{dt}{dz'} dz'
\end{equation}
Reionization of hydrogen is complete by $z=6$ while reionization of helium is not. So, 
$N_0(z=6)$ is the fractional volume density of hydrogen which is $0.93$. Equation 5 can be rewritten as: 
\begin{equation}
N(z=6) = 0.93 + 6.475\times10^{-20}~\int_{z'=z_0}^{z'=6} C~\chi_{{\rm HII}}~\left ( \chi_{{\rm HII}} + \chi_{{\rm HeII}} \right ) ~(1+z')^3~\frac{dt}{dz'} dz'
\end{equation}
Since $\Omega_{b}h^{2}=0.0223$, the comoving baryon
volume density is 0.25 baryons~m$^{-3}$. $\chi_{{\rm HII}}$ and $\chi_{{\rm HeII}}$ are the volume averaged
gas fractions for
ionized hydrogen and ionized helium respectively. Care should be
taken that these are derived with respect to the number density of baryons.
Typical simulations provide ratios with respect to the number density of neutral hydrogen or neutral helium.
In that case, we have adopted $n_{\rm H}$=0.9268n$_{\rm b}$ and $n_{\rm He}$=0.073n$_{\rm b}$.

If we neglect the small contribution to the electron density from Helium, Equation 6
can be simplified to:
\begin{equation}
N(z=6) \approx 0.93 + 6.475\times10^{-20}~\int_{z'=z_0}^{z'=6} C'~(1+z')^3~\frac{dt}{dz'} dz'
\end{equation}
where $C'$ is $<n_{{\rm HII}}^2>/<n_b>^2$, a quantity directly available from the \citet{Trac:07} simulations (See their Figure 12). However, their estimate of $C'$ is an average over the entire comoving volume. In practise, the
clumpiness of gas which is within the source-forming halos is already factored into the estimate of escape fraction.
So, the clumping factor should be a measure of the clumpiness of the IGM excluding halos with star-formation in them (Figure 2a). 
Therefore, we use the exact relation as shown in Equation 6. 

We adopt the most current estimate for the reionization history, redshift dependent
clumping factor (Figure 2a), $n_{e}$, $n_{{\rm HII}}$ and $n_{{\rm HeII}}$
from the high resolution simulations of \citet{Trac:07}. 
A comparison between the number of photons per baryon from these equations
compared with a similar relation derived for complete reionization by \citet{Madau:99} is shown in Figure 2b.
The number of ionizing photons required is sensitive to the reionization history since the recombination rate is proportional
to the square of the density. If reionization is completed earlier, the recombination rate is proportionately 
higher than for a partially ionized medium. In that case, the number of photons per baryon needs to be higher to
maintain the ionized state of the IGM as shown by the triple dot dashed line in Figure 2b. 

We note that the reionization history of \citet{Trac:07} corresponds to the ``late'' reionization
history which would result in a WMAP optical depth to Thomson scattering close to the measured
value of $\tau\sim0.09\pm0.03$ \citep{Spergel:07}. As characterized in \citet{Greif_Bromm} and \citet{Wyithe},
shifting the process of reionization to $z\sim10$ would result in optical depth values which are
at the high end of the observed range.

For each stellar initial mass function that we adopt, we obtain a library of template SED which are
a function of the age of the starburst. We 
first determine the combination of age and mass which results in a best fit to the optical and ultraviolet
luminosity density. This yields the stellar mass density at $z=6$ for that IMF. The range of age and mass
values for a subset of IMFs considered here is shown in Table 1. As emphasized earlier this is a maximal age
for a single stellar population since we adopt a single instantaneous burst.

The Starburst99 model yields the number of hydrogen ionizing photons produced as a function of age of the starburst.
The escape fraction is a measure of the number of ionizing photons escaping from star-forming regions
in galaxies, into the IGM. The escape fraction has been constrained to be
$\sim10-15$\% \citep{Siana:07, Shapley:06} in the Lyman-break galaxy population. 
Thus, we adopt escape fractions of 0.1 and 0.15.
By integrating the number of ionizing photons over the derived stellar age,
we can estimate the total number of photons produced over the
lifetime of the evolving galaxies, multiply that with the escape fraction and compare the result
with the number of ionizing photons required to keep the IGM
ionized as determined from Equation 6.

\subsection{Salpeter IMF}

Figure 3a shows the total number of ionizing photons produced by a starburst for a 
Salpeter IMF which is a power-law of the form dN/dM$\propto$M$^{\alpha}$, $\alpha=-2.3$. The
 metallicity was 0.02 Z$_{\sun}$, escape fraction of 0.1 and clumping factor shown
by the solid black line in Figure 2a. The calculations corresponds to the ``high-V" case
Different lower mass cutoffs were adopted for the Salpeter IMF as shown
by the different colored lines while the upper mass cutoff was fixed at 200\,M$_{\sun}$. The black 
line is the number of ionizing photons required to keep the IGM
ionized between $z<z'<6$. The point where the colored lines which denote the number of ionizing photons
escaping from the star-forming regions cross the black line is the epoch beyond which reionization was complete.

For a Salpeter IMF extending down to 0.1\,M$_{\sun}$ i.e. with no low
mass cutoff, the IGM was ionized for only 60 Myr prior to
$z\sim6$. This is an unphysically small age and indicates that if the 
IMF has a Salpeter shape in $z\sim6$ galaxies, reionization must have been an extremely inhomogeneous process. 

As the low mass cutoff increases from 0.1 to 10 M$_{\sun}$,
the optical luminosity density produced for a fixed age for the starburst decreases. So, as the
low mass cutoff increases, the age of the starburst needs to become younger to reproduce both the UV and optical 
luminosity density. As a result, the number of ionizing photons produced over the duration of a starburst
by a Salpeter IMF with a mass cutoff at 10\,M$_{\sun}$ is lower than that with a mass cutoff at 0.1\,M$_{\sun}$.

Even in the case where the low mass cutoff is at 5 M$_{\sun}$, which produces the largest number of ionizing
photons, the process of reionization has to be shorter than 140 Myr. This indicates that unless
the ratio of clumping factor to escape fraction is significantly lower than what the simulations seem to indicate,
a Salpeter IMF and \zre$>7$ is ruled out. 

The ``low-V" case, since it corresponds to a bluer, stellar population which has a younger age, would result
in fewer ionizing photons over the lifetime of the starburst, making the conclusions
derived above even more stringent. Thus, unless explicitly stated otherwise, our calculations below are done for the
``high-V" case. 

\subsection{Modifications to the High Redshift Stellar IMF}

Motivated by the WMAP results which suggest the possibility of
reionization as early as $z_{reion}\sim11$, we investigate a range
of initial mass functions which would produce sufficient ionizing photons to keep the IGM
ionized at $z>>7$.

As shown in Figure 3a, changing the lower mass cutoff of a Salpeter IMF cannot increase \zre\ above redshift of 7. 
It appears
necessary to change the value of the IMF slope $\alpha$. The flatter the IMF slope,
the larger the contribution of massive stars. Since most of the ionizing photons are produced
by massive stars, even a small increase in the slope results in a significant increase in the
number of ionizing photons produced. 
Changing the low mass cutoff does not change the number of ionizing photons substantially except
for the renormalization by the total mass in the IMF.
If $\alpha=-2$, the IMF has equal mass in logarithmic mass bins. 
So for a value of $\alpha\gtrsim -2$, \zre\ is relatively insensitive to a change in the low mass cutoff
in the IMF. 

Figure 3b shows the number of ionizing photons produced by an IMF which extends from $1-200$\,M$_{\sun}$
and has a range of slopes $\alpha$. For $\alpha=-2.3$, \zre\ falls between $6<z<7$ while for $\alpha=-2.0$,
sufficient ionizing photons are produced to initiate reionization at $z\sim11$ and complete the reionization process
by $z\sim6$.  By comparing Figure 3b and Figure 4, it can be seen that \zre\ for $\alpha\gtrsim -2$
is insensitive to the low mass cutoff of the IMF. The resultant
stellar mass densities at $z\sim6$ for the IMFs considered here are shown in Table 1.

If the formation of high mass stars is suppressed, the number of ionizing photons produced per baryon are reduced.
If the cutoff is at 50 M$_{\sun}$ instead of 200 M$_{\sun}$, the net effect is to increase the 
value of $\alpha$.
For example, a stellar IMF with a mass range of 1-50 M$_{\sun}$, $\alpha=-1.7$ produces a similar number 
of ionizing photons as an IMF extending between 1-200 M$_{\sun}$, $\alpha=-2.1$.

The choice of metallicity in the models is also an important parameter. 
For the calculations we present, the results are based on the 0.02\,Z$_{\sun}$ metallicity models 
which are the lowest available metallicity in Starburst99, although
we also evaluated the results for the 0.4\,Z$_{\sun}$ metallicity templates as illustrated in Figure 3b.
Lower metallicity models 
produce $\sim2-3$ times as many ionizing photons as higher metallicity models (Figure 3b). 
However, we note that the number of ionizing photons produced
for zero metallicity single burst templates which have normal IMFs
as presented by \citet{Schaerer:03}, are very similar to the number of ionizing photons produced for 0.02\,Z$_{\sun}$
metallicity templates to within $\sim$10-20\%. This implies that a Salpeter IMF even with lower metallicity templates than our adopted ones
cannot produce sufficient ionizing photons. Thus, although the exact slope of the derived IMF
is sensitive to the metallicity of the gas collapsing into the stars, the fact that the metallicity enrichment from massive
stars is a relatively quick process and the metallicity of GRB hosts at $z>5$ which are presumably
star-forming galaxies, appear to be $\sim0.01-0.1$\,Z$_{\sun}$, leads us to conclude that
the choice of 0.02\,Z$_{\sun}$ metallicity templates is a reasonable approximation.

\subsection{Dependence between Stellar IMF and \z$_{reion}$}

We have generated template SEDs for a range of stellar initial mass function slopes and stellar mass ranges
which reproduce the observed optical and ultraviolet luminosity density at $z\sim6$. We consider two
cases, one in which the stellar IMF extends between $1-200$\,M$_{\sun}$ and a second in which
the mass range is $1-50$\,M$_{\sun}$. 
As emphasized earlier, the choice of cutoff at the low mass end is insensitive to the result once $\alpha>-2$.
We assume that the metallicity is 0.02\,Z$_{\sun}$ in either case.
We consider a range of $-1>\alpha>-2.6$ and calculate the number of ionizing photons produced for each
scenario. 
We then calculate the redshift at which the number of ionizing photons from the stellar
population exceed the minimum required to maintain the ionized state of hydrogen. It is important to note
that this is not simply a comparison with the solid black lines in Figure 3 and 4 which assume
the particular reionization history (and thereby clumping factor) from the simulations of \citet{Trac:07}.

If reionization were complete at higher redshifts, the recombination rate
between the end of reionization and $z\sim6$ would be higher because of the larger values of $n_{\rm e}$ 
and $n_{\rm HII}$. To estimate this, we simply shift the reionization history and clumping factors of \citet{Trac:07}
to higher redshifts and assume that the clumping factors remained constant once the IGM was completely ionized.
The number of ionizing photons per baryon in the scenario where reionization is complete by $z\sim11$
is plotted as the triple dot dashed line in Figure 2. We note that this is a factor of $\sim$4 higher
than the ``late" reionization history which the models of \citet{Trac:07} provide.
Furthermore, even if the clumping factor were to increase after the IGM was completely reionized,
this would only increase the minimum number of photons per baryon required. Thus, our estimates are,
at worst, a lower limit to the number of photons per baryon required to maintain the ionized state of the IGM. 

We find that
as the redshift of reionization increases, the value of $\alpha$ needs to increase (Figure 5). This is not surprising
since a higher $z_{\rm reion}$ implies a larger number of ionizing photons which are preferentially 
produced in more massive stars.
We also find that as the high mass end of the stellar IMF is cut, from 200 M$_{\sun}$ to 50 M$_{\sun}$,
the slope $\alpha$ needs to increase such that a larger fraction of massive stars make up
 the shortfall in ionizing photons. 

The relation between \zre\ and IMF slope is also sensitive to the ratio of the clumping factor to the escape fraction.
Reduction in the clumping factor decreases the number of ionizing photons required since the recombination
rate is reduced. Similarly, a large escape fraction implies that a larger number of photons escape star-forming regions.
So, a fewer number of ionizing photons need to be produced to reionize the IGM. As a result, in Figure 5, we plot the
relation between \zre\ and $\alpha$ for different ratios of C/f$_{\rm esc}$. A redshift dependent
clumping factor such as that
shown in Figure 2a and an escape fraction of 0.1 is shown as the green lines. The red lines show the estimates
for a clumping factor of 30 and an escape fraction of 0.1. The blue line show the estimates for a clumping factor of 10
and an escape fraction of 0.15 while the cyan line is an extreme case with a clumping factor of 10 and escape fraction of 0.5.
A similar comparison between the redshift when reionization is complete and the IMF slope is shown in Figure 6
for the ``low-V" case.

As can be seen for the ``high-V" case, with a Salpeter IMF ($\alpha=-2.3$), unless the ratio of clumping factor to escape fraction is 
much smaller than 60, the stars in $z\sim6$ galaxies cannot be responsible for reionization at $z>7$.
For our best estimate of the clumping factor and escape fraction, if \zre$=9$, the slope $\alpha=-1.65$ while
if \zre$=11$, $\alpha=-1.5$. In either scenario, only 0.2\% of all baryons at $z\sim6$ are processed
through stars and are responsible for ionizing the IGM.

Our conclusions are therefore broadly consistent with the results of \citet{Tum:04} who argue against the need for a very massive
star population at high redshift.
Their arguments are based on the metal abundance ratios
in metal-poor Galactic halo stars. However, we disagree with them in the sense that
a Salpeter IMF even with a low mass cutoff, is inadequate for producing sufficient ionizing photons to ionize
the IGM while fitting the observed ultraviolet and optical luminosity density at $z\sim6$ (Figure 3). 

Such a low mass turnover in the stellar IMF may indeed be present at $z<2$ as has been argued by
\citet{Dave} and \citet{PvD}. The former attempted to reconcile specific star-formation rate estimates
between models and observations while the latter fit the color evolution of massive early-type cluster 
galaxies at $z<1$. We attempted to fit the observed luminosity densities at $z\sim6$ with the IMF
proposed by \citet{Dave}. We assumed that the $(1+z)^2$ evolution in the mass turnover
of the stellar IMF extends up to $z\sim4$, 
similar to the evolution proposed by \citet{PvD}. Thus, the IMF was assumed to have a slope
of $-1.3$ at $0.1<$M$<12$\,M$_{\sun}$ and $-2.3$ at 200$>$M$>12$\,M$_{\sun}$. For the ``high-V" case, 
I find that this IMF
produces 3.25 ionizing photons per baryon, an age of the stellar population of
90 Myr and a stellar mass of 1.2$\times$10$^7$\,\mun. Unlike the Salpeter IMFs with a sharp cutoff 
shown in Figure 3, such an IMF can account for the the ``late" reionization history of \citet{Trac:07}
and would maintain an ionized IGM at $z\lesssim 7$. However, for the ``low-V" case,
this IMF would produce only 1.0 ionizing photons per baryon, result in a stellar age of 45 Myr and
a stellar mass of 3.7$\times$10$^6$\,\mun, about a factor of 3 lower than the minimum
required to maintain an ionized IGM at $z<7$.

\subsection{An Early Epoch of Reionization from Massive Stars ?}

The fundamental assumption here is that the galaxies seen at $z\sim6$ are the remnants of star-formation responsible
for reionization. It is plausible that the $z\sim6$ galaxies that are seen in deep surveys
are actually comprised of Population II stars with a Salpeter IMF which are responsible for a second late epoch
of reionization. In that scenario, the earlier epochs would be entirely due to Population III
stars which do not contribute to the ultraviolet and visible light luminosity density at $z\sim6$,
a scenario that has been considered previously \citep[e.g][]{Cen:03, Furl:05}. These stars, which are more more massive than 8\,M$_{\sun}$
evolve into black holes within 30 Myr, i.e. by $z\sim6$.
The number of ionizing photons required to keep
the IGM ionized between $15<z<6$ is $\sim$11 photons/baryon. An IMF extending between
$10-100$\,M$_{\sun}$, with a Salpeter slope of $2.3$ provides a total of 1.38 photons per baryon for a stellar
mass density of 2.5$\times$10$^{6}$\,\mun. The stellar mass density in massive Population III stars must therefore
be 2$\times$10$^{7}$~\mun. Thus, if
reionization was initiated by massive stars which evolve into neutron stars and black holes by $z\sim6$,
there must be as much mass density in these remnants as in the stars that we detect in the galaxies.

If in a contrived scenario,
the initial $z\sim13$ epoch of reionization from Population III stars was relatively brief,
lasting $\Delta z=2$ \citep[e.g.][]{Cen:03}, it would require  
that the stellar mass density in the initial burst was $\sim$10$^{6}$\,\mun.
This corresponds to 10\% of baryons that are in stars at $z\sim6$.
Since the lifetime of 10\,M$_{\sun}$ stars
are $\sim$30 Myr while the interval between $13<z<15$ spans 60 Myr, the first epoch of reionization must be
relatively inhomogeneous and will depend on the exact
epoch at which star-formation in dark matter halos was initiated. 
Furthermore, the truncation of Population III star-formation through feedback processes argues against
them being significant contributors to reionization \citep{Greif_Bromm}. Measuring the
distribution of Stromgren sphere size using high spatial resolution HI observations will reveal the true nature
of the reionization history at $z>10$.

It should be noted that Population III star formation could potentially extend down to
lower redshifts, depending on the effect of feedback on the metallicity of 
the star-forming environments. The rates of star-formation
in such stars
are however thought to be $3\times10^{-4}$ of the Population II star-formation rate and are inconsequential to
the ultraviolet luminosity density or the co-moving star-formation rate density \citep{Tornatore, Brook:07}.

\section{Conclusions}

I have utilized existing data from deep {\it Hubble} and {\it Spitzer} surveys to provide a measure
of the co-moving luminosity density at ultraviolet and optical wavelengths at $z\sim6$. In particular,
I provide a measure of the contribution to the optical luminosity density from faint galaxies which
are below the {\it Spitzer} detection limit. Even after accounting for faint galaxies, the optical
luminosity density is a factor of $2-3$ below the ultraviolet luminosity density. I fit the resultant
luminosity density estimates with a stellar population synthesis model to determine the maximal age and
stellar mass density at $z\sim6$. Assuming a Salpeter IMF, I find that the stellar mass density at $z\sim6$
is 1.6$\times$10$^{7}$\,\mun and the maximal age of a single stellar population is $\lesssim$100 Myr. By comparing
the number of ionizing photons per baryon produced over the age of the starburst with the number of ionizing photons
per baryon
required to keep the IGM ionized, I find that reionization must have been a brief
inhomogeneous process lasting $\lesssim$100 Myr and must have been completed as late as $z<7$ if the
stellar IMF had a Salpeter slope. 

Motivated by WMAP results which suggest an early epoch of reionization,
I investigate the form of the stellar IMF if reionization was a single continuous process at redshifts higher than 6.
If the past history of star-formation in $z\sim6$ galaxies was responsible for reionization,
I find that the the slope of the stellar IMF has to be non-Salpeter with a slope of $\alpha=-1.65$ if \zre=9
and $\alpha=-1.5$ if \zre=11 for an IMF extending up to 200 M$_{\sun}$. However, the exact
slope is sensitive to the ratio of the clumping factor to escape fraction, the metallicity of the stars
and the value of the visible luminosity density at $z\sim6$. 

On the other hand, the IGM could have been
ionized between $6<z<15$ by a population of massive stars between $10-100$\,M$_{\sun}$
which evolve into stellar
remnants like black holes by $z\sim6$. Such a population would not be constrained by the rest-frame visible
and ultraviolet luminosity density at $z\sim6$. The resultant comoving mass density of remnants at $z\sim6$ would
be $2\times10^{7}$\,\mun\ and would therefore be comparable to the stellar mass density that we see in $z\sim6$ galaxies.

Finally, I consider the evolving stellar initial mass function that has been suggested by \citet{Dave}
and \citet{PvD} to account for the discrepancy in specific star-formation rate estimates and colors
of early-type galaxies at lower redshifts
respectively. This IMF if evolved out to $z\sim4$, would
result in a bottom-light IMF with a non-Salpeter slope at $M<12$\,M$_{\sun}$. Such an IMF could produce
sufficient ionizing photons to account for late reionization of the IGM at $z\lesssim7$ 
only if the visible light luminosity density was at the high end of our assumed range.

{\it Acknowledgments:}
I would like to acknowledge Avi Loeb for stimulating discussions and guidance.
I would also like to thank Hy Trac for providing tabulated estimates of the ionized fraction and clumping factor
from his simulations and for helpful suggestions. 
I am very grateful to Rachel Somerville for providing early results from her semi-analytical
models. 
This research is partially supported by the Spitzer Space Telescope Theoretical Research Program which
was provided by NASA through a contract issued by the Jet Propulsion Laboratory, California Institute of Technology.

\begin{deluxetable}{lcccc}
\tabletypesize{\scriptsize}
\tablecaption{Stellar Mass Density at $z\sim6$ for Different IMFs}
\tablewidth{0pt}

\tablehead{
\colhead{Mass Range} &
\colhead{IMF Slope} &
\colhead{Stellar Mass Density} &
\colhead{Stellar Age\tablenotemark{a}} &
\colhead{Ionizing photons per baryon\tablenotemark{b}}\\

\colhead{M$_{\sun}$} &
\colhead{$\alpha$} &
\colhead{10$^{6}$\,\mun} &
\colhead{Myr} &
\colhead{}
}
\startdata
$0.1-200$ & $-2.3$	&	16.1 & 95 & 0.91 \\
$1-200$ & $-2.3$	&	 7.3 & 96 & 2.06\\
$2-200$ & $-2.3$	&	 6.1 & 100 & 2.69 \\
$5-200$ & $-2.3$	&	 5.1 & 68  & 3.97 \\
$10-200$ & $-2.3$	&	 2.5 & 25 & 5.50 \\\\

$1-200$ & $-2.0$	&	15.1 & 93 & 3.36 \\
$1-200$ & $-1.7$	&	 9.5 & 94 & 4.58 \\
$0.5-200$ & $-2.0$	&	10.7 & 94 & 2.96 \\
$0.5-200$ & $-1.7$	&	15.6 & 92 & 4.36 \\
$1-50$ & $-2.0$	&	 7.0 & 95 & 1.88\\
$1-50$ & $-1.7$	&	 8.6 & 93 & 2.77\\
$1-50$ & $-1.4$	&	 11.6 & 91 & 3.72 \\
$1-50$ & $-1.0$	&	 19.0 & 88 & 4.88 \\
\enddata
\tablenotetext{a}{This is only a maximal age for a single stellar population
assuming all the stars formed in a single instantaneous burst and were
then allowed to passively evolve. The 
exact age is sensitive to the e-folding timescale of a starburst and cannot be derived with the data at hand.}
\tablenotetext{b}{This is the total number of ionizing photons produced over the duration of the starburst
per baryon per 10$^{6}$ M$_{\sun}$ of stars formed. This needs to be multiplied by the escape fraction
before a comparison with the recombination rate is made.}
\end{deluxetable}

\begin{deluxetable}{lcc}
\tabletypesize{\scriptsize}
\tablecaption{Two Component Stellar Mass Density at $z\sim6$ for Different IMFs}
\tablewidth{0pt}

\tablehead{
\colhead{Mass Range} &
\colhead{IMF Slope} &
\colhead{Maximal Stellar Mass Density\tablenotemark{c}} \\

\colhead{M$_{\sun}$} &
\colhead{$\alpha$} &
\colhead{10$^{6}$\,\mun} 
}
\startdata
$0.1-200$ & $-2.3$      & 46 \\
$1-200$ & $-2.3$        & 21\\
$2-200$ & $-2.3$        & 47 \\
$5-200$ & $-2.3$        & ... \\
$10-200$ & $-2.3$       & ...\\\\

$1-200$ & $-2.0$        & 36 \\
$1-200$ & $-1.7$        & 73 \\
$0.5-200$ & $-2.0$      & 39\\
$0.5-200$ & $-1.7$      & 74\\
$1-50$ & $-2.0$        & 26\\
$1-50$ & $-1.7$        & 42 \\
$1-50$ & $-1.4$        & 73 \\
$1-50$ & $-1.0$        & 176\\
\enddata
\tablenotetext{c}{Maximal stellar mass density estimates at $z\sim6$ derived using a two component
stellar population. The young stellar component has an age of 10 Myr and dominates
the ultraviolet luminosity density. The older stellar component has an age of 770 Myr and dominates
the stellar mass density. See text for details.}
\end{deluxetable}

\begin{figure}
\plottwo{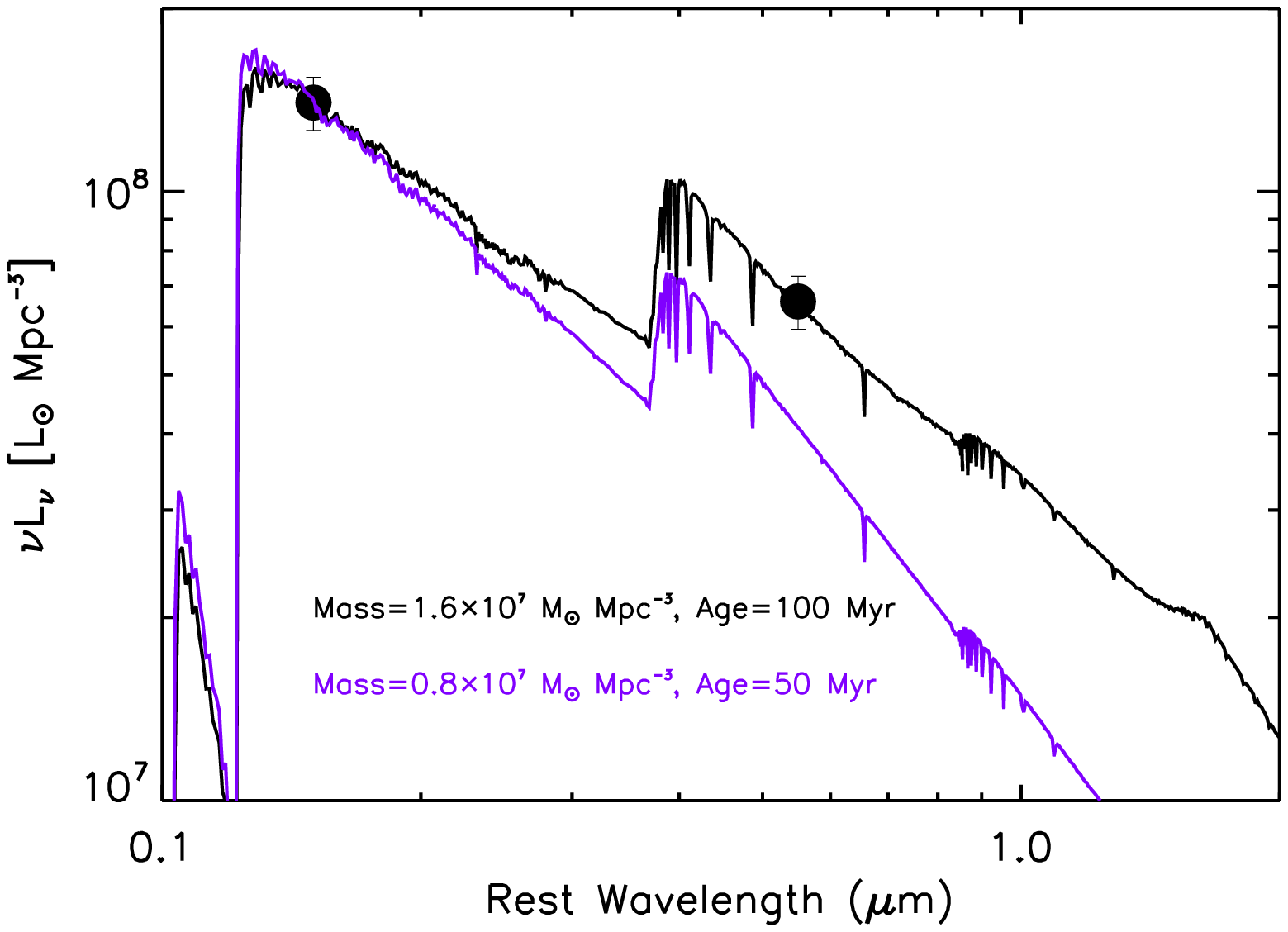}{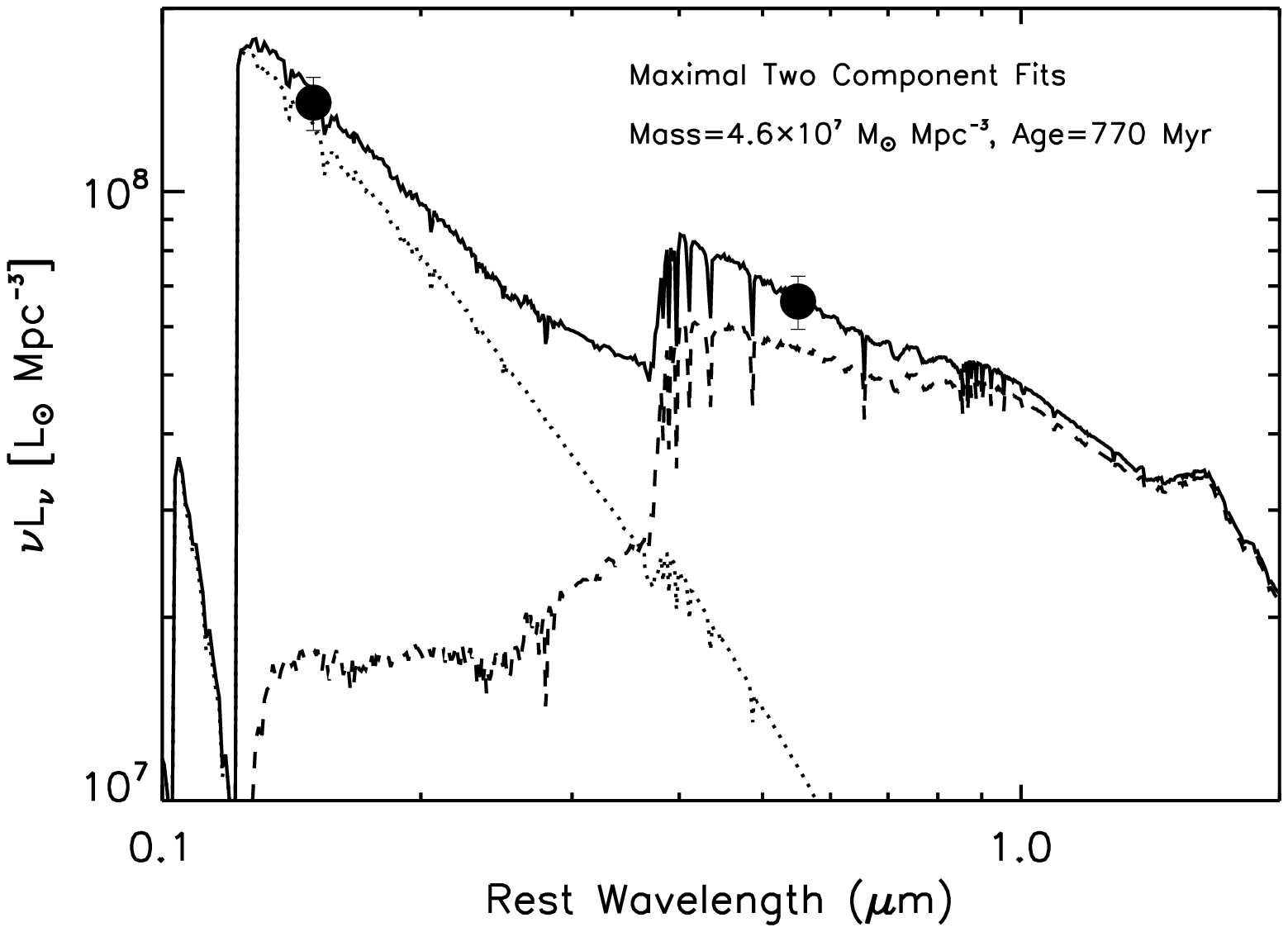}
\caption{Left Panel: Solid circles show the rest-frame 1500\AA\ and 5500\AA\ luminosity density at $z\sim6$ after appropriate completeness corrections for
faint galaxies have been made. The faint galaxies have been assumed to have visible/ultraviolet luminosity ratios similar
to that of $z\sim6$ galaxies detected in {\it Spitzer}/GOODS observations (the ``high-V" case).
The solid black line shows the age and stellar mass density of the template SED which fits the luminosity densities
for a Salpeter stellar IMF with no dust extinction.
The age indicates that if the stellar mass function had been Salpeter, the stellar population that is seen in $z\sim6$
galaxies must have formed as late as 100 Myr earlier i.e. at $z\sim6.5$. If faint galaxies have lower visible to ultraviolet flux ratios compared to bright galaxies, the $V-$band luminosity density would be decreased by 30\% (the ``low-V" case).
The purple line is the spectral energy distribution which fits the luminosity density values in this 
latter scenario. Right Panel: Maximal fits to the ``high-V" luminosity density at $z\sim6$ assuming two stellar components and a Salpeter IMF. The UV luminosity density is fit by a 10 Myr old burst while a 770 Myr old,
high mass to light ratio stellar component, is
the dominant contributor to the visible luminosity density.
}
\end{figure}

\begin{figure}
\plottwo{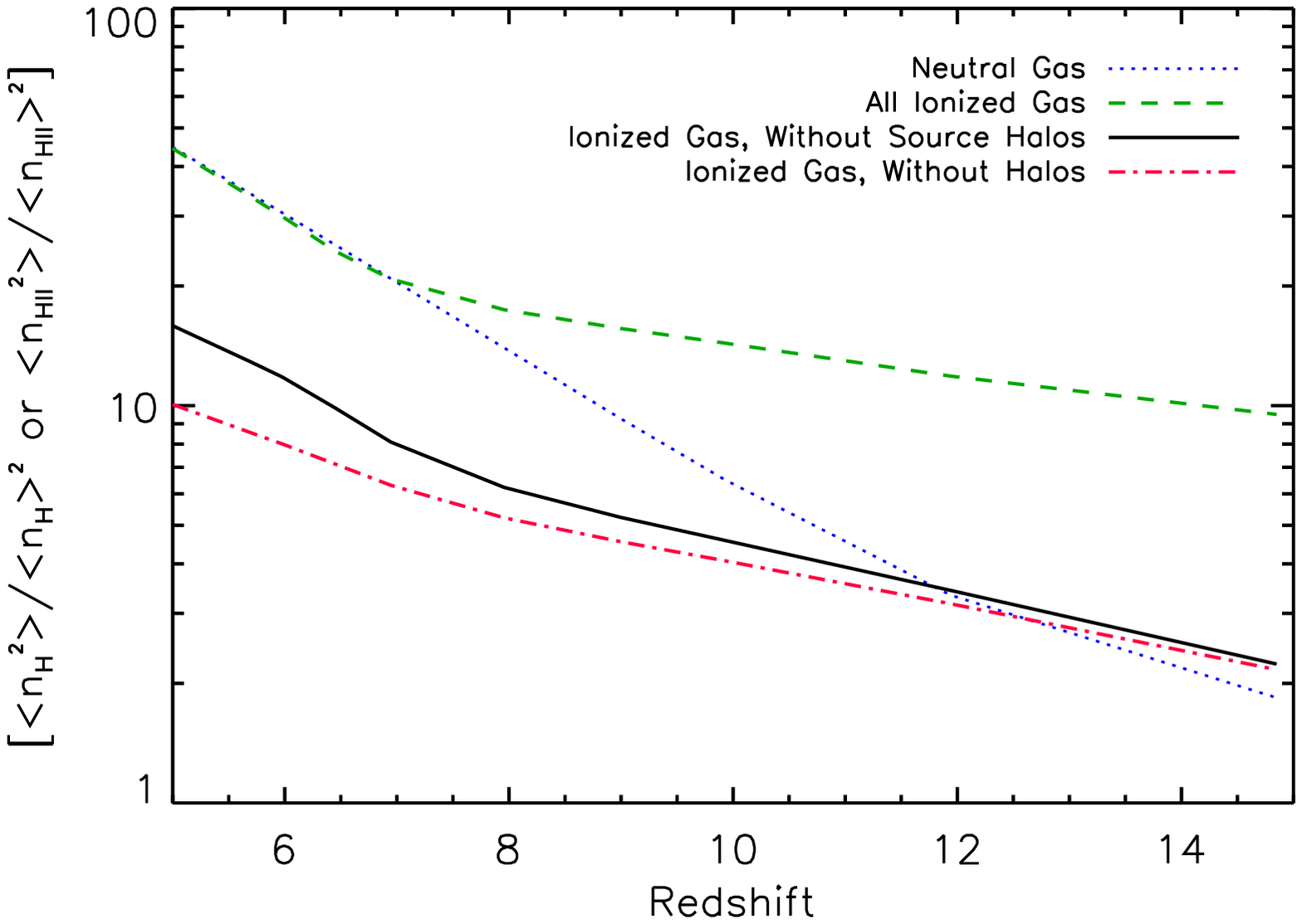}{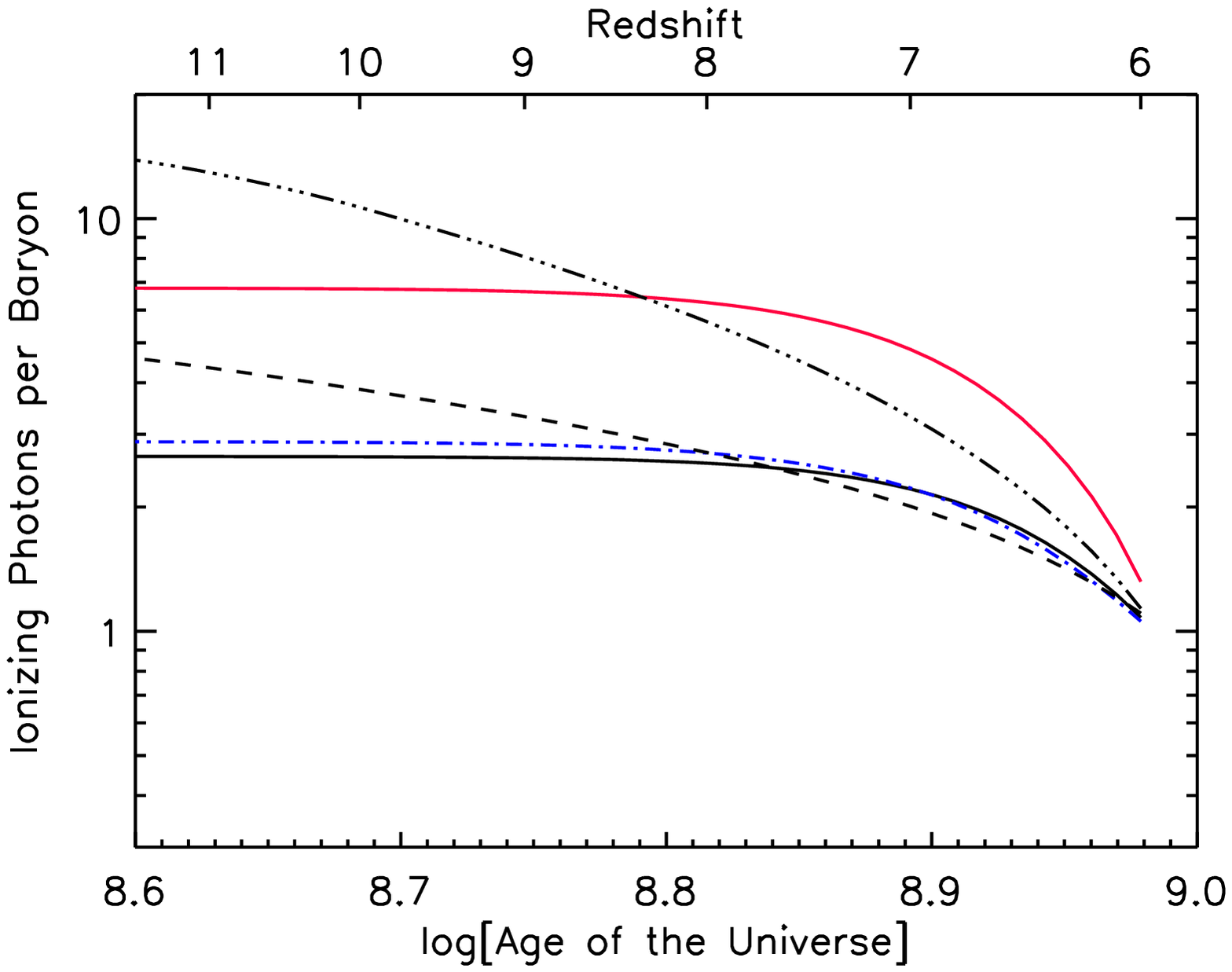}
\caption{
{\it Left Panel:} Comparison between the clumping factors of neutral and ionized gas from the \citet{Trac:07}
simulations for their reionization history. The blue dotted line shows the clumping of neutral gas averaged over the simulation volume.
The green dashed line is the clumping of ionized gas averaged over the simulation volume.
The solid black line is the clumping of ionized gas after star-forming halos have been excluded.
The dot-dash red line is the clumping of ionized gas after all halos are excluded. Since, the clumpiness
of gas in the star-forming halos is already factored into the escape fraction, we use the solid black line
as the best estimate of the clumping factor.
{\it Right Panel:} The plot shows the number of ionizing photons per baryon required to ionize the IGM and account for recombinations. 
The solid black line is the number of ionizing photons per baryon assuming the reionization history and redshift dependent
clumping factor of \citet{Trac:07} after source halos have been excluded (solid black line in the left
panel). The triple dot dashed line assumes that reionization was complete by $z\sim12$ and the clumping
factor and escape fraction remained constant at lower redshifts.
The solid red 
line assumes the same reionization history but uses a constant
clumping factor of 30. The dot-dash blue line assumes a constant clumping factor of 10. 
The dashed line is the 
number of photons derived
using the parameteric solution of \citet{Madau:99} for complete hydrogen and helium reionization
and the redshift dependent clumping factor of \citet{Trac:07}.
The reionization history and properties of the IGM
are clearly important in determining the number of ionizing photons required.
}
\end{figure}

\begin{figure}
\plottwo{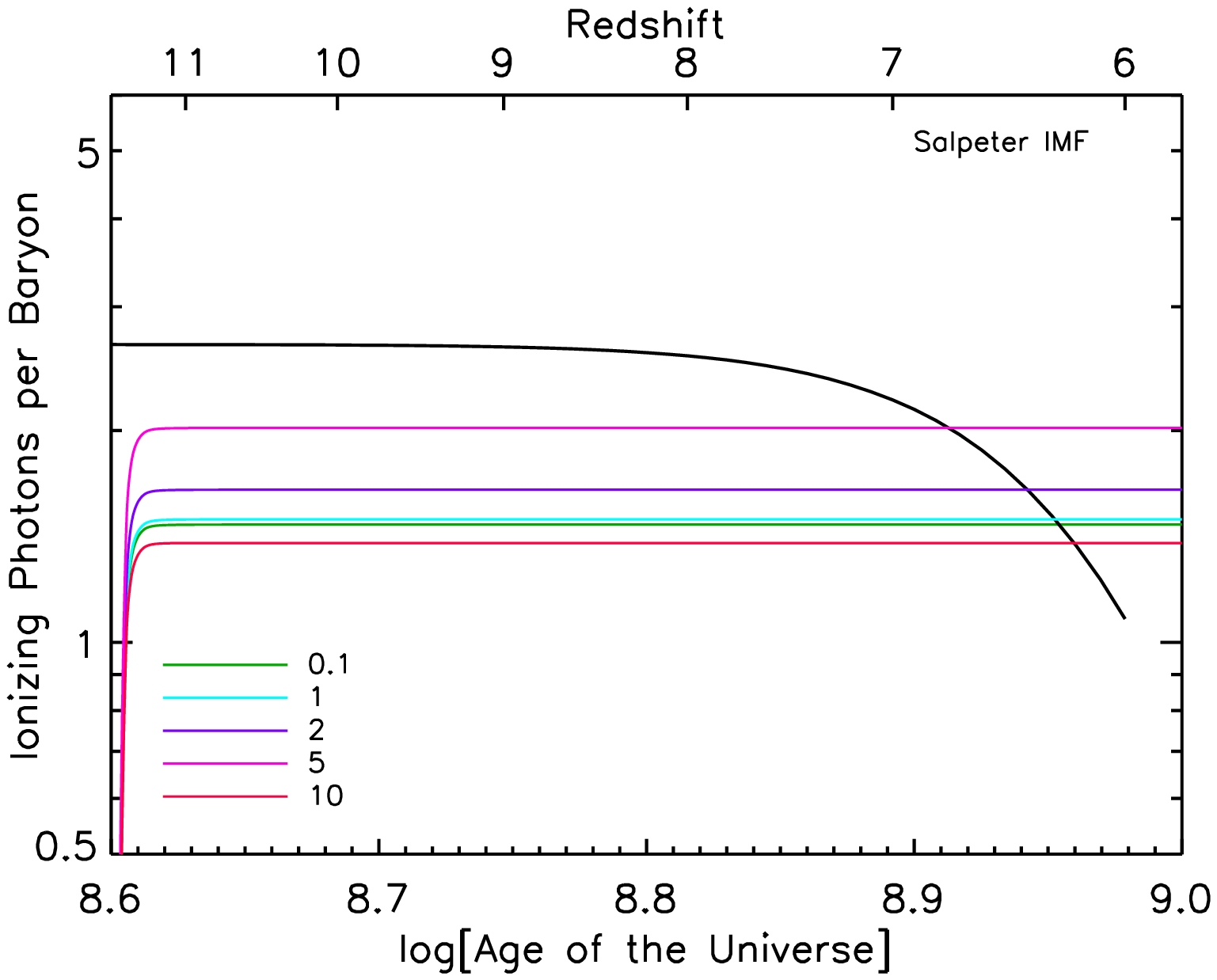}{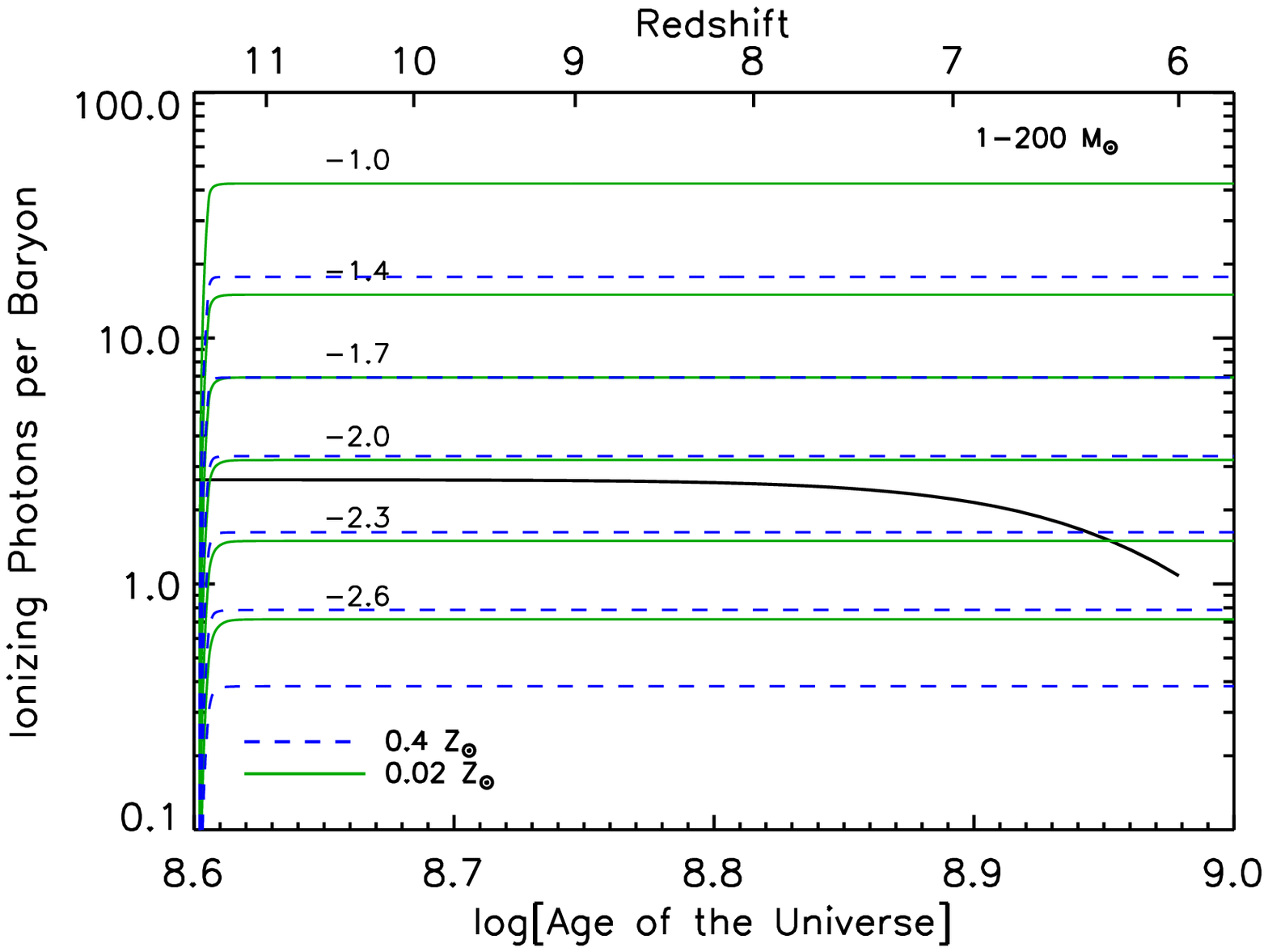}
\caption{{\it Left Panel:} The solid black line shows the minimum number of ionizing photons per baryon required to 
ionize the IGM and account for recombinations
between the plotted redshift and $z\sim6$ (Same as Figure 2). Overplotted as solid lines of different colors
are the number of ionizing photons per baryon
produced for a 0.02 Z$_{\sun}$ Salpeter IMF (dN/dM$\propto$M$^{-2.3}$) which reproduces
the observed ultraviolet and optical luminosity density at $z\sim6$. The different colors
correspond to different low mass cutoffs; 0.1, 1, 2, 5, 10 M$_{\sun}$.
The stellar mass
density at $z\sim6$ corresponding to these different IMFs are shown in Table 1.
Reionization would be complete from the time at which the colored lines cross the solid black line.
For the assumed clumping factor and escape fraction, a Salpeter IMF, even with a low mass cutoff,
would maintain the ionized state of the IGM only at $z<7$.
{\it Right Panel:} The solid green lines show the number of ionizing photons for a 0.02 Z$_{\sun}$
non-Salpeter IMF,
with stellar masses spanning 1-200 M$_{\sun}$ and different values of $\alpha$, the IMF slope. The dashed blue line
shows the number of ionizing photons for a 0.4 Z$_{\sun}$ model template with the same range of slopes. Clearly, metallicity
is an important parameter in constraining the IMF. 
}
\end{figure}

\begin{figure}
\plotone{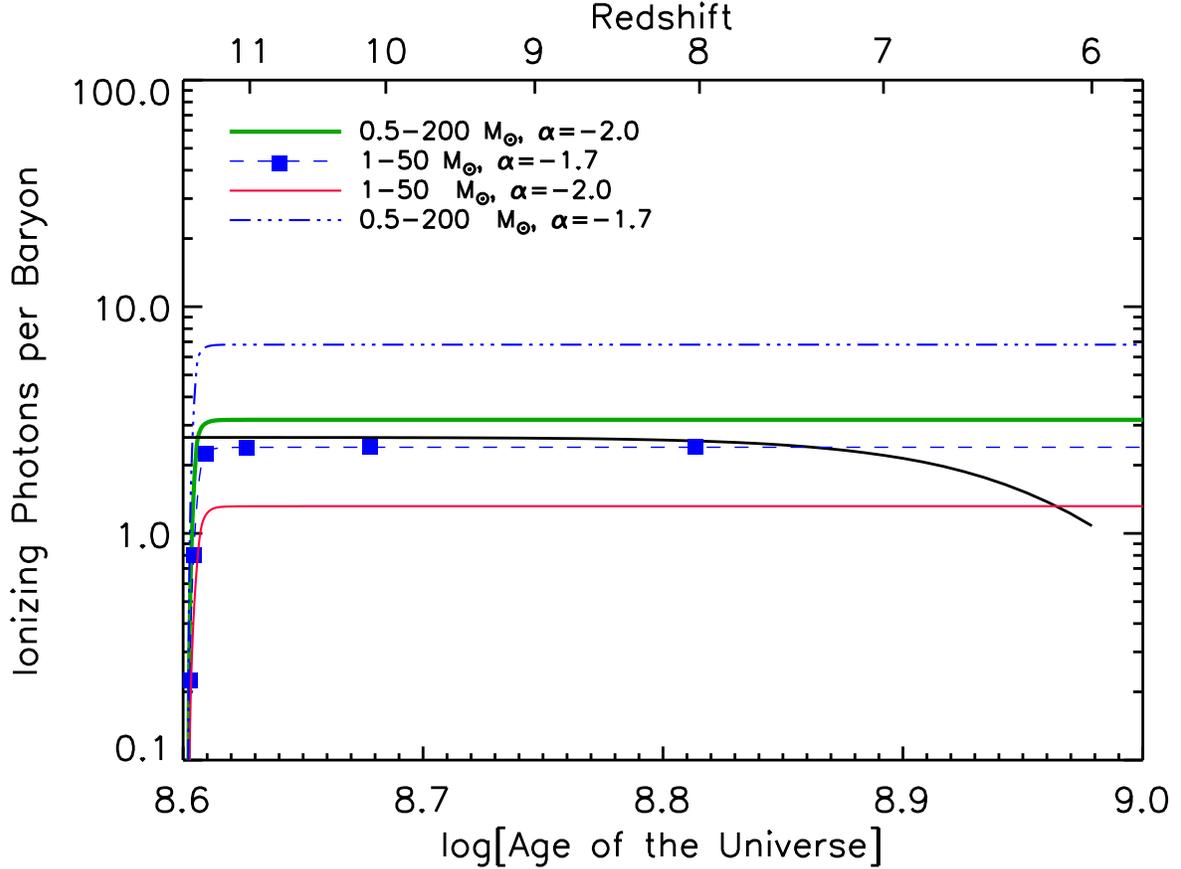}
\caption{
Dependence between stellar IMF and number of ionizing photons produced, as a function of IMF parameters for
the reionization history of \citet{Trac:07}. For the ``late" reionization history considered by \citet{Trac:07},
either the ratio of clumping factor to escape fraction must be much lower than the best estimates
or the stellar IMF needs to have
a slope index of $\alpha\sim-2.1$ if the IMF extends up to 200\,M$_{\sun}$. 
If the IMF has a high mass cutoff at $\sim$50\,M$_{\sun}$, the slope needs to be $\alpha\sim-1.65$.
In either scenario, this is substantially flatter than a Salpeter slope ($\alpha$=$-2.3$).
}
\end{figure}

\begin{figure}
\plotone{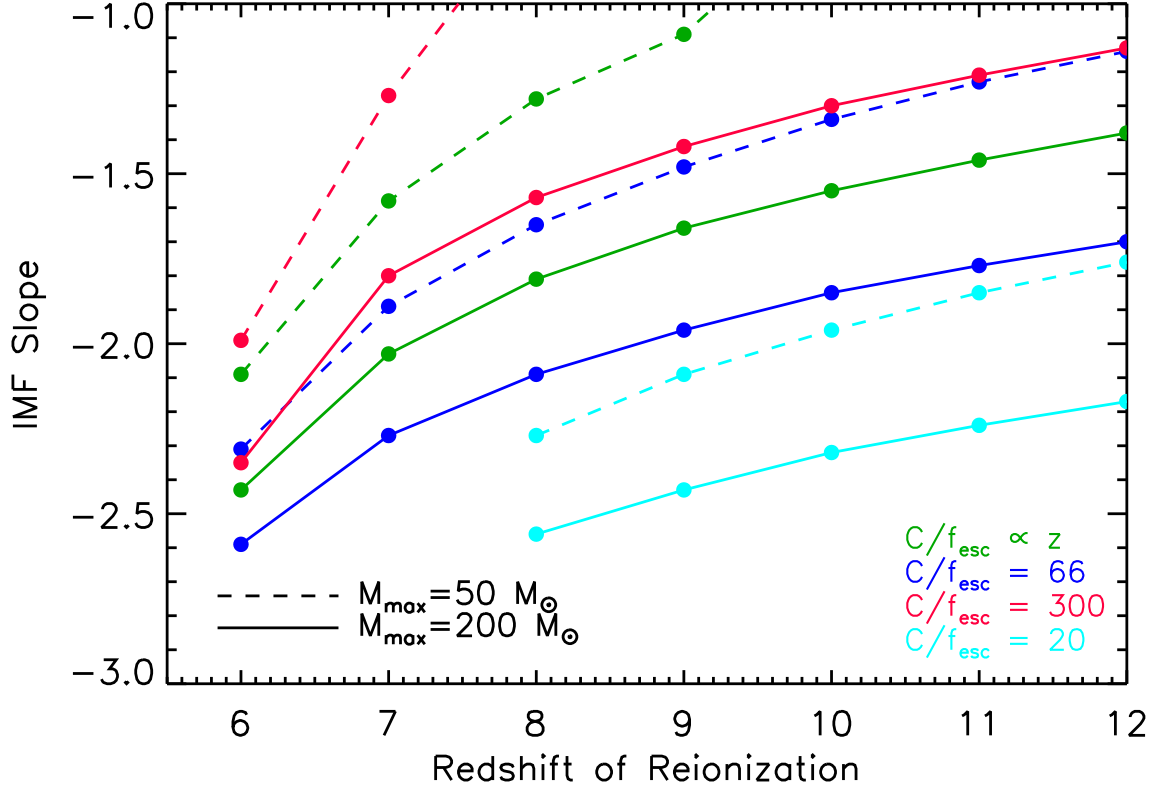}
\caption{
Dependence between redshift of reionization and slope of the stellar initial mass function for the ``high-V"
case.
The solid lines show the case where the IMF extends between $1-200$~M$_{\sun}$ while the
dashed lines have the IMF extending between $1-50$~M$_{\sun}$. Four different values of C/f$_{\rm esc}$
are shown. The solid green lines are our best estimate which
assumes the redshift dependent clumping factor derived by the
simulations of \citet{Trac:07} and an escape fraction of 0.1. The blue lines assume a
redshift independent clumping factor of 10 and an escape fraction of 0.15. The red lines assume a constant
clumping factor of 30 and escape fraction of 0.1 while the cyan lines are an extreme case
assuming a clumping factor of 10 and an escape fraction of 0.5. 
For our best estimate of the clumping factor and escape fraction, 
the IMF must have a slope dN/dM$\propto{\rm M}^{-1.65}$ for reionization to 
be complete by $z\sim9$ and dN/dM$\propto{\rm M}^{-1.5}$ for reionization to be complete by $z\sim11$.
}
\end{figure}

\begin{figure}
\plotone{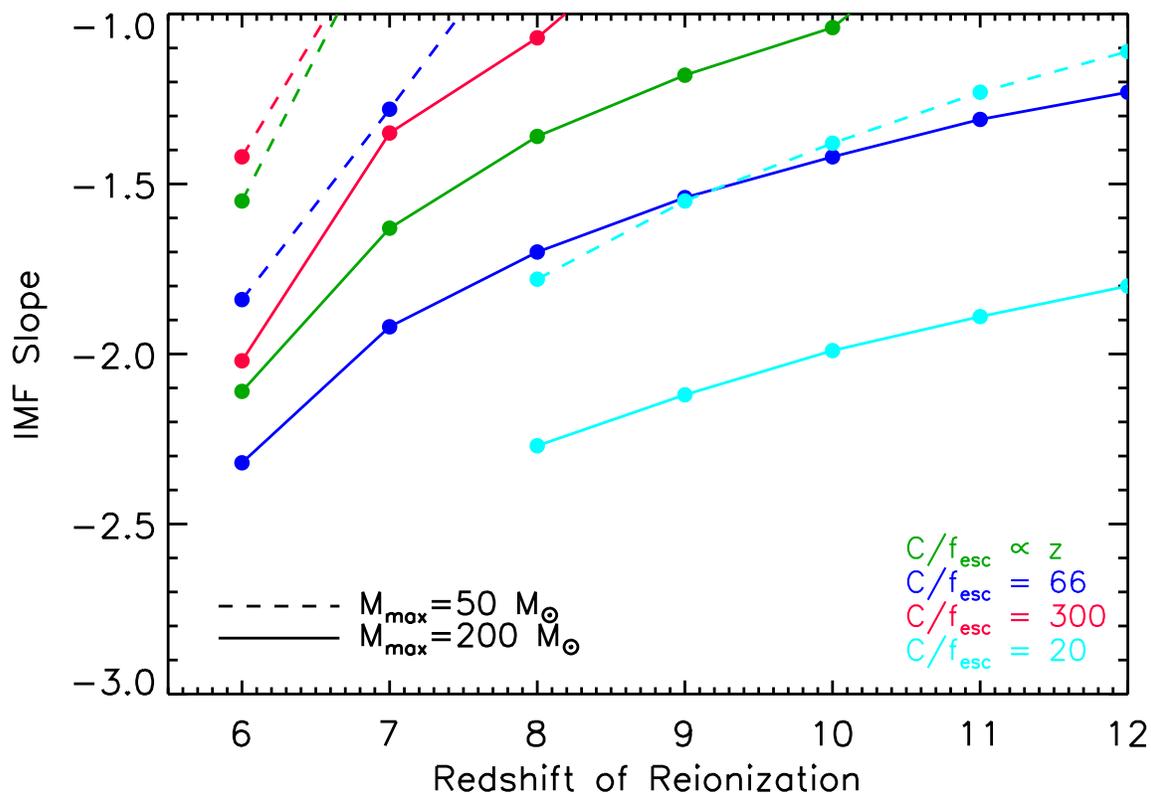}
\caption{
Similar to Figure 5 for the ``low-V" case where the V-band luminosity density is 30\% lower
as shown in Figure 1. Since a lower V-band luminosity density implies that fewer stars
have been produced, the slope of the stellar IMF has to be even flatter resulting in even more high mass stars
compared to Figure 5.
}
\end{figure}

\end{document}